\documentclass[12pt]{iopart}
\usepackage{graphicx}
\usepackage{braket}
\usepackage{hyperref}
\usepackage{amssymb}
\usepackage{tabularx}
\begin{document}

\title[Temperature dependence of cluster decay]{Temperature dependence of cluster decay}

\author{D. F. Rojas-Gamboa and N. G. Kelkar}

\address{Departamento de Fisica, Universidad de los Andes,
Carrera 1 No. 18 A - 10, Bogot\'a 111711, Colombia}
\eads{\mailto{df.rojas11@uniandes.edu.co}, \mailto{nkelkar@uniandes.edu.co}}
\author{O. L. Caballero}

\address{Department of Physics, University of Guelph, Guelph,
ON N1G 2W1, Canada}
\eads{\mailto{ocaballe@uoguelph.ca}}
\vspace{10pt}
\begin{indented}
\item[]July 2022
\end{indented}

\begin{abstract}
A universal decay law (UDL) for light cluster decay of excited nuclei is formulated by fitting the UDL half-lives to those evaluated within an excitation energy-dependent double folding model (DFM). The half-lives are evaluated within a preformed cluster model. The excitation energy dependence is introduced both in the energy of the emitted cluster and in the interaction potential through the density distributions of the interacting nuclei. The half-lives are found to decrease when the difference between the excitation energies of the parent and daughter nuclei, $\Delta E^{*} > $ 0, increases, with the reduction being a few orders of magnitude for higher values of $\Delta E^{*}$. This can be of importance for nucleosynthesis calculations of heavy elements formed in extremely hot environments, as well as in highly energetic heavy-ion collisions. The UDL for excited nuclei is used to provide an estimate of the enhancement in the cluster decay rates of thermally excited nuclei such as those produced in r-process nucleosynthesis. 
\end{abstract}

\vspace{2pc}
\noindent{\it Keywords}: Cluster decay, thermal excitations, r-process nucleosynthesis, hot nuclei.
%
%
%

\section{Introduction}

In the investigation of heavy elements, the half-lives and their possible decay modes test the ideas developed on the nuclear structure. This has motivated the formulation of semiempirical relations that provide nuclear information that would have otherwise been difficult to obtain through a complicated microscopic framework. For alpha ($\alpha$) decay, the Geiger-Nuttall law \cite{GeigerNuttall1911} proves to be one of the most versatile tools for studying half-life systematics. This was extended by Viola and Seaborg \cite{Viola1966}, incorporating a dependence on the parent proton number $Z$. Different expressions have been tried to generalize the Geiger-Nuttall law to cluster decay \cite{Royer2000, Royer2010}. The half-lives of nuclei decaying by the emission of a wide range of charged particles has been explained by a universal decay law (UDL) in \cite{Qi-etal2009}. Here we formulate a UDL to study the emission of light nuclei, namely, $^{14}$C, $^{20}$O, $^{23}$F, $^{24-26}$Ne, $^{28,30}$Mg, and $^{32,34}$Si from parent nuclei ranging between mass numbers of 222 to 242, i.e., from  $^{222}$Ra to $^{242}$Cm. The latter phenomenon is known as cluster radioactivity or cluster decay; predicted in 1980 \cite{Sandulescu1980} and first observed in the emission of $^{14}$C from $^{223}$Ra \cite{Rose1984}.

Within a semiclassical framework, the spontaneous emission of a charged particle (or light nucleus) can be described as a quantum tunneling phenomenon of the particle through the Coulomb barrier. This requires the assumption of a preformed cluster of the emitted nuclei inside the parent nucleus. The half-life or the decay width are then simply expressed in terms of the penetration probability $P$, the assault frequency $\nu$, and the cluster preformation factor $P_c$ as
\begin{equation}\label{eq:halflife}
    t_{1/2}=\frac{\ln2}{\lambda_0}=\frac{\ln2}{P_c\,\nu\,P} \,,
\end{equation}
where $\lambda_0$ is the decay constant related to the width as, $\hbar \lambda_0 = \Gamma$. The appearance of a preformation probability $P_c$ in (\ref{eq:halflife}) takes into account the non-zero probability for the existence of the preformed cluster inside the parent nucleus \cite{Xu-etal2016,Xu-etal2017} and is closely related to the internal structure of nuclei.

The study of nuclear properties at finite temperatures is relevant both for applications in astrophysics \cite{Lattimer,BetheRevModPhys,Botvina} and for models of finite nuclei and nuclear matter at high excitation energy \cite{Benvenuto,Morrisey,Shlomo1991}. On the other hand, the internal excitations of nuclei play an important role in regulating their abundance, since they increase significantly their entropy. The excitations form an important ingredient in multifragmentation studies of hot nuclei. For example, the authors in \cite{BotvinaPLB}, working within a statistical multifragmentation model find significant temperature-dependent modifications relevant for stellar dynamics and nucleosynthesis.  They perform calculations for supernova matter by assuming that the nuclei have the same internal temperature as the surrounding medium.

The present work as mentioned above is focused on determining the excitation energy (related to nuclear temperature) dependence and thereby the ambient temperature dependence of cluster decay half-lives from the UDL and is organized as follows. We present the extension of the commonly used double folding model (DFM) to take into account the possible excitations of the parent as well as the heavy daughter nuclei. The excitation energy dependence in the DFM is introduced through temperature-dependent matter and charge density distributions of the daughter nuclei. The $Q$ value or the energy of the emitted light cluster is also modified in order to take the excitations into account. The modified DFM is used to calculate the cluster decay half-lives within the tunneling formalism. An excitation energy-dependent universal decay law (UDL) is formulated by fitting the parameters to reproduce the excitation energy-dependent half-lives evaluated using the DFM. Finally, the excitation energy-dependent UDL is used to calculate the ambient temperature-dependent half-lives. The results display a reduction in half-lives (or enhanced decay rates) with increasing ambient temperature, which can be of relevance at high temperatures such as those encountered in r-process nucleosynthesis producing the heavy elements.


\section{Formalism for cluster decay} \label{formalism}
In the absence of sufficient information on the cluster decay rates of excited nuclei to daughters in their ground or excited states, we begin with a theoretical model which reproduces the measured cluster decay half-lives well and then extend it within reasonable theoretical assumptions to those for excited parent and daughter nuclei. The latter exercise is carried out by relating the excitation energy of the nucleus to a nuclear temperature and formulating a double folding model (DFM) with temperature-dependent nuclear densities. The excitation energy of the parent nucleus is incorporated through an effective $Q$ value which reflects as a shift in the energy of the tunneling light cluster which is taken to be in its ground state. Results of this ``microscopic calculation'' are then used to fit a universal decay law dependent on the excitation energies of the parent and daughter nuclei.

\subsection{Microscopic calculation}
Cluster decay half-lives (\ref{eq:halflife}) in literature are usually calculated using the expression of the decay width $\Gamma$ considering the emission of a charged particle (light nucleus, here referred to as cluster) in the transition of the parent nucleus in its ground state to the ground state of the daughter (i.e., $g.s.\rightarrow g.s.$) \cite{Gurvitz1987,Castaneda2007, Kelkar2016}. We refer to the heavier decay product as the daughter nucleus. The light cluster is also considered to be in its ground state. To extend the cluster decay half-life calculations to excited states, the excitation energy relative to the ground state is included in both the energy of the emitted cluster and the interaction potential. Therefore, the energy of the cluster is taken as an effective $Q$ value given by
\begin{equation}\label{eq:Q_eff}
    Q_{eff}\equiv Q+E_p^{*}-E_d^{*}=Q+\Delta E^{*}\,.
\end{equation}
where $E_p^{*}$ and $E_d^{*}$ are the excitation energies of the parent and daughter nuclei, respectively, and $\Delta E^{*}$ is the difference between them. Since the cluster decay is treated as a tunneling problem in which the preformed cluster tunnels the potential barrier composed of the nuclear, Coulomb, and centrifugal potentials, the definitions of these are of relevance in the construction of the excitation energy-dependent interaction potential and will hence be explained in some detail below.

The nuclear potential is calculated using a realistic nucleon-nucleon (NN) interaction folded with the density distributions of both interacting nuclei (i.e. the heavy daughter and the light cluster which are preformed in the parent) \cite{Castaneda2007,Kelkar2016}. The excitation energy effects are included via the nuclear density distributions defining a nuclear temperature $T$ of the daughter nucleus as \cite{Lang1959}
\begin{equation}\label{eq:Eexc_temp}
    E_d^{*}(T) = \frac{A}{9}T^2\,.
\end{equation}
The temperature-dependent matter density distribution of the excited daughter nucleus is given as \cite{Antonov1989,GuptaSinghGreiner},
\begin{equation}\label{eq:density_temp}
    \rho_d\left(r;E_d^*(T)\right)=\rho_{0}(T)\left[1+\exp \left(\frac{r-R(T)} 
{\tilde{a}(T)}\right)\right]^{-1}\,,
\end{equation}
where $\rho_{0}(T)$ is obtained by normalizing the density to the mass number, $\int\rho(r,T)d\mathbf{r}=A$, and $R(T)$ and $\tilde{a}(T)$ are given by
\begin{equation}\label{eq:Rtemp}
    R(T)=R_{0}\left(1+0.0005\,T^{2}\right)
\end{equation}
and
\begin{equation}\label{eq:atemp}
    \tilde{a}(T)=\tilde{a}_0\left(1+0.01\,T^{2}\right)\,
\end{equation}
where $R_{0}=1.07A^{1/3}$ fm and $\tilde{a}_0=0.54$ fm. The nuclear potential is then obtained within the double-folding model \cite{Satchler1979} as
\begin{equation}\label{eq:V_Double-Folding}
    V_{\rm N}\left(\mathbf{r};E_d^*(T)\right)=\int \mathrm{d} \mathbf{r}_{c} \mathrm{d} \mathbf{r}_{d}\,\rho_{c}\left(\mathbf{r}_{c}\right) v_{\rm N}\left(|\mathbf{s}|=\left|\mathbf{r}+\mathbf{r}_{c}-\mathbf{r}_{d}\right|\right) \rho_{d}\left(\mathbf{r}_{d};E_d^*(T)\right)
\end{equation}
where a popular choice of the effective NN interaction based on the M3Y-Reid-type soft core potential is used. Thus,  
\begin{equation}
    v_{\rm N}\left(|\mathbf{s}|\right)=7999 \frac{\exp \left(-4\left|\mathbf{s}\right|\right)}{4\left|\mathbf{s}\right|}-2134 \frac{\exp \left(-2.5\left|\mathbf{s}\right|\right)}{2.5\left|\mathbf{s}\right|} + J_{00}\delta(\mathbf{s})\,,
\end{equation}
where $|\mathbf{s}|=\left|\mathbf{r}+\mathbf{r}_{c}-\mathbf{r}_{d}\right|$ is the distance between a nucleon in the daughter nucleus and a nucleon in the cluster. This consists of a short-ranged repulsion and a long-ranged attraction responsible for the direct component of the interaction. An additional zero-range contribution with $J_{00}=-276(1-0.005\,E/A_c)$ called the knock-on exchange term takes into account the antisymmetrization of identical nucleons in the cluster and the daughter nucleus. The latter represents a kind of nonlocality in the double folding potential \cite{Adel2017}. Nonlocality can be included in the description of nuclear decay using other approaches too and we refer the reader to \cite{Jhoan2019,Diego2022} for further reading.

The excitation energy-dependent Coulomb potential is evaluated in a similar manner by using the $T$-dependent charge density distributions as in Eqs. (\ref{eq:density_temp}) normalized to the atomic number $Z$. Thus, the total potential is given by, 
\begin{equation}\label{eq:potential_temp}
    V\left(r;E_p^{*},E_d^{*}\right) = \lambda\left(E_p^{*},E_d^{*}\right)\, V_{\rm N}\left(r;E_d^{*}\right)+V_{\rm C}\left(r;E_d^{*}\right) +\frac{\hbar^{2}}{2\mu}\frac{\left(l+\frac{1}{2}\right)^{2}}{r^{2}}\,.
\end{equation}
where $r$ is the separation between the center of masses of the cluster and the daughter nucleus. Note that the usual centrifugal barrier has been replaced by the Langer modified one \cite{Langer1937} since the width is evaluated within a semiclassical approximation. $\lambda\left(E_p^{*},E_d^{*}\right)$ is the strength of the nuclear interaction and is fixed by imposing the Bohr-Sommerfeld quantization condition. This also determines the depth of the nuclear potential for different excited states of the parent and daughter nuclei as
\begin{equation}\label{eq:BS_temp}
    \int_{r_{1}(E_p^{*},E_d^{*})}^{r_{2}(E_p^{*},E_d^{*})} k\left(r;E_p^{*},E_d^{*}\right) \,{\rm d r}=(G-l+1) \frac{\pi}{2}
\end{equation}
where $k\left(r;E_p^{*},E_d^{*}\right)=\sqrt{\frac{2\mu}{\hbar^2}\left|V\left(r;E_p^{*},E_d^{*}\right)-Q_{eff}\right|}$ is the wave number. The turning points depend parametrically on the excited states through the condition $V\left(r_i; E_p^{*}, E_d^{*}\right) = Q_{eff}$, and $n$ is the number of nodes of the quasibound wave function of the light cluster - daughter nucleus relative motion. This is expressed as $n = (G-l)/2$, where $G$ is a global quantum number and $l$ is the relative orbital angular momentum quantum number. The global quantum number $G$ is chosen according to the Wildermuth-Tang condition \cite{Wildermuth2013, Ni-Ren2010},
\begin{equation}\label{eq:Wildermuth_rule}
    G=2 n+l=\sum_{i=1}^{A_{c}}\left(g_{i}^{(A_{c}+A_{d})}-g_{i}^{(A_{c})}\right) \,,
\end{equation}
where $g_i^{(Ad+Ac)}$ are the oscillator quantum numbers of the nucleons forming the cluster required to ensure that the cluster is  completely outside the shell occupied by the daughter nucleus, and $g_i^{(Ac)}$ are the internal quantum numbers of the nucleons in the emitted cluster. The values of $g_i$ are taken as $g_i=4$ for nucleons in the $50\leq Z$, $N\leq82$ shell, $g_i=5$ for nucleons in the $82 < Z$, $N\leq126$ shell, $g_i=6$ for nucleons in the $N\leq184$, and $g_i=7$ for nucleons outside the $N=184$ neutron shell closure. These values correspond to the $4\hbar\omega$, $5\hbar\omega$, $6\hbar\omega$, and $7\hbar\omega$ harmonic-oscillator shells, respectively. The Wildermuth-Tang condition is considered to be sufficient to account for the main effects of the Pauli principle.

As a consequence of the excitation energy dependence of the interaction potential, and the effective $Q$ value, the decay width for a preformed cluster model within the semiclassical Jeffreys-Wentzel–Kramers–Brillouin (JWKB) approximation \cite{Gurvitz1987,Castaneda2007} becomes excitation energy-dependent and is given by
\begin{eqnarray}\label{eq:Gamma_temp}
    \Gamma\left(E_p^{*},E_d^{*}\right) = P_c\,\frac{\hbar^{2}}{2 \mu}&\left[\int_{r_{1}(E_p^{*},E_d^{*})}^{r_{2}(E_p^{*},E_d^{*})} \frac{\rm d r}{k\left(r;E_p^{*},E_d^{*}\right)}\right]^{-1}\nonumber\\
    &\times\exp\left[-2 \int_{r_{1}(E_p^{*},E_d^{*})}^{r_{2}(E_p^{*},E_d^{*})} k\left(r;E_p^{*},E_d^{*}\right) \,\rm d r\right]\,.
\end{eqnarray}
where $P_c$ is the cluster preformation probability and $\mu$ is the reduced mass of the two nuclei forming the cluster inside the parent nucleus. The exponential factor corresponds to the penetration probability and the factor in front of it is the normalization of the bound-state wave function in the region between $r_1$ and $r_2$, the first and second turning points. The excitation energy-dependent half-life is then calculated as $t_{1/2}\left(E_p^{*},E_d^{*}\right) = \hbar\,\ln2\,/\,\Gamma\left(E_p^{*},E_d^{*}\right)$.


\subsection{Universal decay law for excited nuclei}\label{UniversalDecayLaw}
The relation between the decay constant and the penetration probability, the assault frequency, and the preformation factor allows us to write a logarithmic formula for the half-life (\ref{eq:halflife}) which is convenient for obtaining a parameterized analytical formula for the latter. Thus, we begin by writing, 
\begin{equation}\label{eq:UDL_T0}
    \log{(\nu \,t_{1/2})}=-\log{P}-\log{P_c} + {\rm constant}. 
\end{equation}

In the simplest model, one can write the first term $-\log{P}$ assuming a tunneling process in which the light cluster goes through the Coulomb barrier formed due to its interaction with the daughter nucleus, with the nuclear potential being a rectangular well of width $R_0=R_c+R_d$. The penetration probability is given by
\begin{equation}\label{eq:penetration_prob}
    P = \exp\left[-2 \int_{R_{0}}^{R} k(r) \,\rm d r\right]\,,
\end{equation}
where $k\left(r\right)=\sqrt{\frac{2\mu}{\hbar^2}\left|V(r)-E\right|}$ is the wavenumber inside the barrier and $R$ is the outer classical turning point calculated from $V_C(R)=E$. Here, $E$ is the energy of the tunneling particle and $\mu$ is the reduced mass of the cluster-daughter nucleus system. The interaction potential $V(r)$ has contributions from both the Coulomb and centrifugal potentials, i.e.,
\begin{equation}
    V(r) = Z_cZ_d\frac{e^2}{r} + \frac{\hbar^2}{2\mu}\frac{l(l+1)}{r^2}\,.
\end{equation}
 
Thus, the wavenumber is given by $k(r)=\sqrt{\frac{2\mu E}{\hbar^2}\left[\left(\frac{R}{r}-1\right)+\sigma\frac{1}{r^2}\right]}$, where $\sigma=\frac{\hbar^2}{2\mu E}l(l+1)$ is a small quantity due to the centrifugal potential being very small compared with the Coulomb potential \cite{Gamow1949,Zhang-etal2011}. With these considerations, the integral in equation (\ref{eq:penetration_prob}) is calculated as
\begin{eqnarray}\label{eq:int_k-l_temp}
    \fl\int_{R_0}^{R}k(r)\,dr=\sqrt{\frac{2\mu E}{\hbar^2}}R&\left[\cos^{-1}\left(\frac{R_0}{R}\right)^{1/2}-\left(\frac{R_0}{R}\right)^{1/2}\left(1-\frac{R_0}{R}\right)^{1/2}\right]\nonumber\\
    &+\sigma\sqrt{\frac{2\mu E}{\hbar^2}}\frac{1}{\sqrt{R_0R}}\sqrt{1-\frac{R_0}{R}}\,.
\end{eqnarray}

This result is in good agreement with that obtained in \cite{Qi-etal2009, Delion2009} in which the penetrability is written in terms of Coulomb-Hankel functions. To simplify \ref{eq:int_k-l_temp}, we consider the potential barrier is relatively wide, i.e., $R>>R_0$. This allows us to make the approximations $\cos^{-1}\approx\frac{\pi}{2}-\frac{R_0}{R}$ and $\left(1-\frac{R_0}{R}\right)^{1/2}\approx1$. The integral of equation (\ref{eq:int_k-l_temp}) is then approximated as
\begin{eqnarray}\label{eq:int_k-l}
    \int_{R_0}^{R}k(r)\,dr=\sqrt{\frac{2\mu E}{\hbar^2}}R\left[\frac{\pi}{2}-2\left(\frac{R_0}{R}\right)^{1/2}\right]+\sigma\sqrt{\frac{2\mu E}{\hbar^2}}\frac{1}{\sqrt{R_0R}}\,.
\end{eqnarray}

Replacing (\ref{eq:int_k-l}) in (\ref{eq:penetration_prob}), choosing the energy of the emitted cluster $E$ to be $Q_{eff}=Q+\Delta E^*$ as in (\ref{eq:Q_eff}), and defining the turning points by $E=Z_cZ_d\,e^2/R$ and $R_0=r_0\left(A_c^{1/3}+A_d^{1/3}\right)$, the penetration probability (\ref{eq:penetration_prob}) is then given as
\begin{equation}\label{eq:logP-l-depen_temp}
    \log{P(T)} = -\alpha\left(E_p^{*},E_d^{*}\right)\,\chi'-b\,\rho'-c\,\frac{l(l+1)}{\rho'}
\end{equation}
where the coefficients are $\alpha\left(E_p^{*},E_d^{*}\right)=\left(a_0-a_1\frac{\Delta E^{*}}{Q}\right)$, with $a_0=\frac{\pi e^{2}\sqrt{2m_0}}{\hbar\ln 10}$ and $a_1=a_0/2$, $b=-\frac{4e\sqrt{2m_{0}r_{0}}}{\hbar\ln 10}$, and $c=-\frac{2\hbar}{\ln10 e\sqrt{2m_0r_0}}$, and the functions $\chi'$ and $\rho'$ are defined as 

\begin{equation*}
 \chi'=Z_cZ_d\sqrt{\frac{A_cA_d}{A_pE}}
\end{equation*}
and 
\begin{equation*}
 \rho'= \sqrt{Z_cZ_d\frac{A_cA_d}{A_p}\left(A_c^{1/3}+A_d^{1/3}\right)}.
\end{equation*}

We note that the first two terms of the UDL (\ref{eq:logP-l-depen_temp}) are the same as in Ref. \cite{Qi-etal2009} for $E_p^{*}=E_d^{*}=0$ and arise from the Coulomb interaction. The third term introduces a correction associated with the angular momentum carried by the emitted cluster \cite{Qi-etal2012}. 

The contribution of the assault frequency to the UDL is taken to be the same as that in  Refs. \cite{Dong-etal2010, Zhang2011}. The cluster is assumed to vibrate near the surface of the parent nucleus in a harmonic oscillator potential with frequency $\omega$ and reduced mass $\mu$. Using the virial theorem $\mu\omega^{2}\braket{r^{2}} = \left(2n_{r}+l+{3}/{2}\right)\hbar\omega$ where the (radial) number of nodes $n_r$ and angular momentum $l$ are related to the global quantum number by $G = 2n_{r}+l$, and the rms radius can be written in terms of the radius of the parent nucleus as $\braket{r^2} = (3/5)R_p^2$. Assuming masses are proportional to the mass number and replacing $R_i = r_0 A_i^{1/3}$, the assault frequency $\nu = \omega/2\pi$ is then given as \cite{Dong-etal2010, Zhang2011}
\begin{equation}\label{eq:log-nu} 
    \nu = \left[\left(G+\frac{3}{2}\right)\frac{A_p^{1/3}}{A_cA_d}\right]\nu_0\,,
\end{equation}
where $\nu_0 = \hbar/(1.2\pi m_0r_0^2)\approx1.4714\times10^{22}$ s$^{-1}$. Some authors \cite{Denisov2009,Akrawy2018,Soylu2021} also include the nuclear isospin asymmetry term $I=(N_p-Z_p)/A_p$ for the parent nucleus and the parity effects. Using the above inputs, we formulate the universal decay law (UDL) for the excitation energy-dependent half-life as
\begin{eqnarray}\label{eq:half-life_temp}
    \fl\log{[\nu_0 \,t_{1/2}^{\rm{(UDL)}}\left(E_p^{*},E_d^{*}\right)]} = &\alpha\left(E_p^{*},E_d^{*}\right)\,\chi'+ b\,\rho' + c
    \,\frac{l(l+1)}{\rho'}\nonumber\\
    &+d\,\log\left[\left(G+\frac{3}{2}\right)\frac{A_p^{1/3}}{A_cA_d}\right]+
    e\,I(I+1)+f\left(1-(-1)^l\right)\,,
\end{eqnarray}
where, $\alpha\left(E_p^{*},E_d^{*}\right)=\left(a_0-a_1\frac{\Delta E^{*}}{Q}\right)$. Though $a_0$, $a_1$, $b$, $c$, $d$, $e$, and $f$ are constants as defined above, they can be considered as free parameters to be fitted in order to make up for the approximations made in deriving the UDL. An important feature emerges from the above excitation energy-dependent UDL. Since the correction introduced by the angular momentum is very small, as mentioned above, the half-life can be factorized into two terms in which one of them depends just on the half-life for the (parent) ground state to the (daughter) ground state decay and the other that is excitation-energy-dependent. Thus, 
\begin{equation}\label{eq:half-life_temp_factorized}
    \log\left[\nu_0\,t_{1/2}\left(E_p^{*},E_d^{*}\right)\right] = \log\left[\nu_0\,t_{1/2}(g.s.\rightarrow g.s.)\right] - a_1\,\chi'\,\frac{\Delta E^{*}}{Q}\,.
\end{equation}
where $t_{1/2}(g.s.\rightarrow g.s.)$ is the half-life when the emission occurs from ground state to ground state. We can therefore fit the single parameter $a_1$ by evaluating the ratio $t_{1/2}(E_p^{*},E_d^{*}) / t_{1/2}(g.s.\rightarrow g.s.)$ using the Eq. (\ref{eq:Gamma_temp}) for the width in the double folding model and  $t_{1/2}\left(E_p^{*},E_d^{*}\right) = \hbar\,\ln2\,/\,\Gamma\left(E_p^{*},E_d^{*}\right)$. Since it is the ratio which we are fitting, we do not need to specify the preformation probability $P_c$ as long we assume that it remains the same for nuclei in the ground and excited states. $P_c$ in the DFM is generally obtained phenomenologically by comparing the theoretical half-lives with experimental data. Having thus fitted the parameter $a_1$ (to be discussed in the next section), we can now consider  (\ref{eq:half-life_temp_factorized}) to be the universal decay law for excited nuclei. The first term in (\ref{eq:half-life_temp_factorized}) can be replaced by the experimental half-life (if available) or using a UDL such those in Refs. \cite{Akrawy2018,Soylu2021,Ismail2022}. Note that the objective of this study is not to fit the universal decay law for the g.s. $\to$ g.s. decays but rather to obtain an expression which can be used for excited nuclei. Hence, it suffices for us to fit the parameter $a_1$ and use a UDL available from literature \cite{Akrawy2018,Soylu2021,Ismail2022} for the g.s. $\to$ g.s. decays.


\section{Results and discussions}\label{results}
The UDL is a simple relation in radioactive decay relating the half-life and basic information of the parent nucleus, the daughter nucleus, and the emitted cluster, such as mass and atomic numbers, $Q$-value, and the angular momentum. In the present work, we formulate a UDL for the decay of ground as well as excited states of parent nuclei to the ground and excited states of the daughter nuclei. The latter is done partly with the objective of evaluating the half-lives of thermally excited nuclei in an astrophysical environment at high temperatures, $T_s$, of the environment, such as those encountered in the r-process nucleosynthesis sites. Since the evaluation of a temperature-dependent half-life for each nucleus within a model calculation is a cumbersome task, a UDL which depends on temperature and the basic information mentioned above can prove useful in testing the influence of the thermally enhanced decay rates in nucleosynthesis calculations. 

\subsection{Excitation energy dependent half-lives}
To investigate the effect of the excitation energy in cluster decay, we calculate numerically the half-lives of some even-even and odd-$A$ nuclei using the DFM (with $\Gamma$ evaluated as in (\ref{eq:Gamma_temp}) assuming $P_c=1$). The excitation energy dependence as mentioned earlier is introduced through both the energy of the emitted cluster and the charge and matter density distribution via the half-density radius, (\ref{eq:Rtemp}), and the surface diffuseness, (\ref{eq:atemp}). The inclusion of excitation energy leads to a shift of the second and third turning points that are solutions of $V\left(r;E_p^{*},E_d^{*}\right)=Q_{eff}$, where the total potential is obtained using the folding procedure, Eqs. (\ref{eq:V_Double-Folding}), (\ref{eq:density_temp}), and (\ref{eq:potential_temp}), with the strength of the nuclear potential $\lambda$ fixed at each combination of excitation energies of the parent and daughter nuclei by the Bohr-Sommerfeld quantization condition, (\ref{eq:BS_temp}). Even though the strength of the nuclear part gets modified when $E_p^{*}$ and $E_d^{*}$ change, the net effect is dominated by the modification of the width of the Coulomb barrier, which leads to a change in the penetration probability. The shift in the turning points with excitation energy is not large but the penetration probability being an exponential factor is sensitive to this change and grows rapidly with decreasing width of the barrier. Therefore, the cluster decay half-lives are affected by the excitation energy of the parent and daughter nuclei.

\begin{figure}[h!]
    \centering
    \includegraphics[scale=0.9]{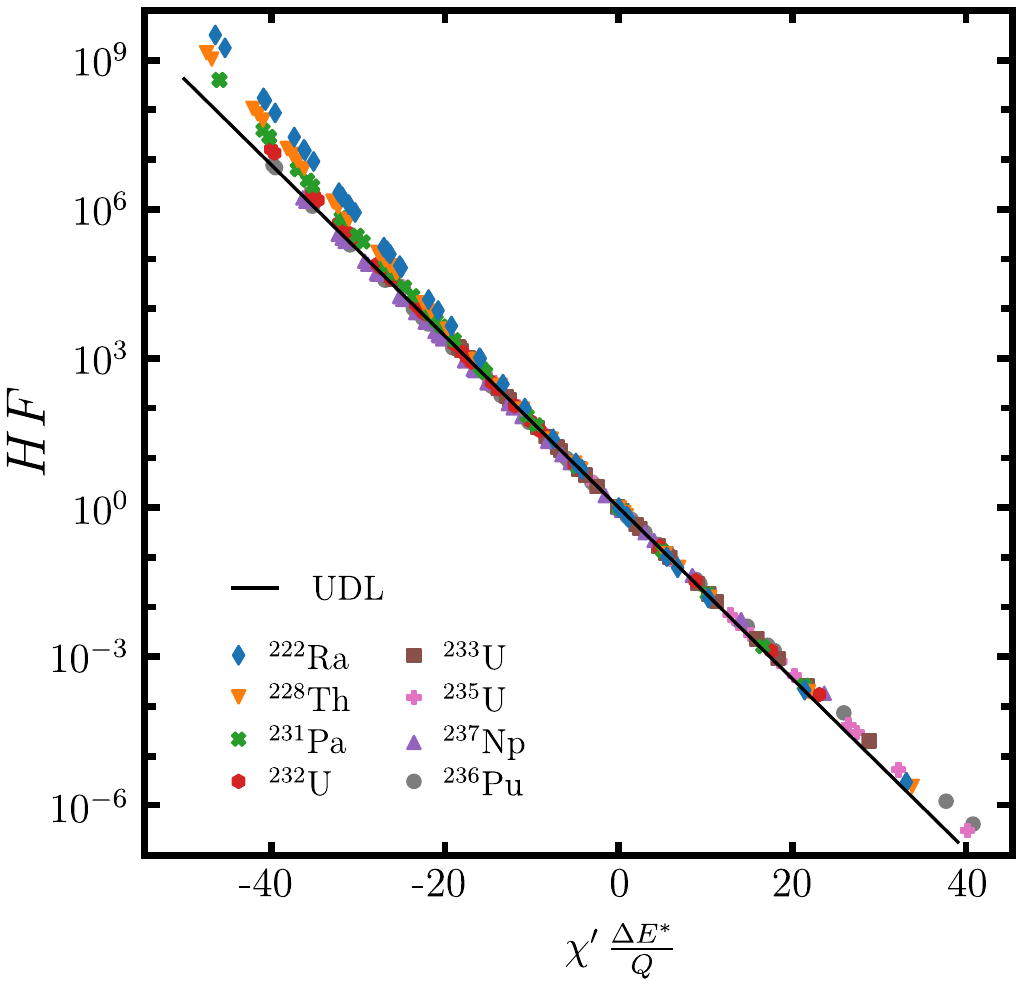}
    \caption{Hindrance factor ($HF$) in semilog scale for cluster emission from heavy nuclei for different values of $E_p^*$ and $E_d^*$.}
    \label{fig:halflife_exc-level}
\end{figure}

The excitation energy-dependent UDL is fitted with the results obtained from the DFM for cluster decay half-lives of cluster emitters decaying from their ground and excited states to the ground and excited states of the daughter nuclei. Since the excitation energy-dependent UDL can be factorized into a zero-temperature part plus an excitation energy-dependent one (\ref{eq:half-life_temp_factorized}), it is possible to calculate the coefficients of the latter separately from those with the ground state to ground state emission. To this end, eight heavy nuclei, including even-even and odd-A, are selected. For each decay, the decaying parent nucleus is in any state of a set of six of its energy levels including the ground state. Similarly, the daughter nucleus is in any state of a set of six states, including its ground state. Thus, for each decay, 36 transitions result from all the possible combinations for parent and daughter states. We limit the calculations to six excited states each of the parent and heavy daughter since considering higher levels result in very small tunneling energies of the light cluster nucleus. The latter would lead to insignificant tunneling probabilities. Apart from this we must also mention that the light cluster nucleus is chosen always to be in its ground state since the first excited state of the light nuclei considered here is of the order of few MeV and less likely to be reached by a thermal excitation as compared to the keV levels of the heavier parent and daughter nuclei. Using the nuclear data from National Nuclear Data Center \cite{NNDC}, the parameter estimation of $a_1$ within standard \textit{curve fit} Python library is $0.1724 \pm 0.0006$ MeV$^{1/2}$.

A hindrance factor ($HF$) could be defined as the ratio of the half-lives for emissions in which the parent nucleus or the daughter nucleus or both nuclei are excited and those in which the emission occurs from the ground state to ground state,
\begin{equation}\label{eq:HF}
    HF = \frac{t_{1/2}\left(E_p^*,E_d^*\right)}{t_{1/2}(g.s.\rightarrow g.s.)} = 10^{-a_1\,\chi'\,\Delta E^{*} / Q}\,.
\end{equation}
The logarithm of $HF$ is proportional to the difference in the excitation energies of the parent and daughter nuclei $\Delta E^*$ \cite{Delion2009}, which is considered a relevant nuclear property of the excited states' half-lives. Figure (\ref{fig:halflife_exc-level}) displays the $HF$ for different cluster emitters. This figure shows the UDL (\ref{eq:half-life_temp_factorized}), or the $HF$ (\ref{eq:HF}), is a good approximation for the decays involving $\left|\chi'\frac{\Delta E^{*}}{Q}\right|\lesssim 20$ MeV$^{-1/2}$, where the daughter nucleus is in the ground state or low-lying excited states. These results also suggest an enhancement in the cluster radioactivity process when the excitation energy of the parent nucleus is larger than that of the daughter nucleus.

\subsection{Temperature-dependent half-lives}
Nucleosynthesis involving the production of heavy and super-heavy nuclei is expected to happen at extremely high temperatures under explosive conditions \cite{BotvinaPLB}. Numerical calculations of the abundance of heavy elements such as in the $r$-process nucleosynthesis require the inputs from different nuclear reactions as well as decay processes. The light cluster decay mode is usually neglected in the $r$-process abundance calculations because the probability of occurrence is usually lower compared to that of other decay modes. Here, we explore the possibility of the cluster decay channel becoming important at elevated temperatures of the environment.
\begin{figure}[h!]
    \centering
    \includegraphics[scale=0.9]{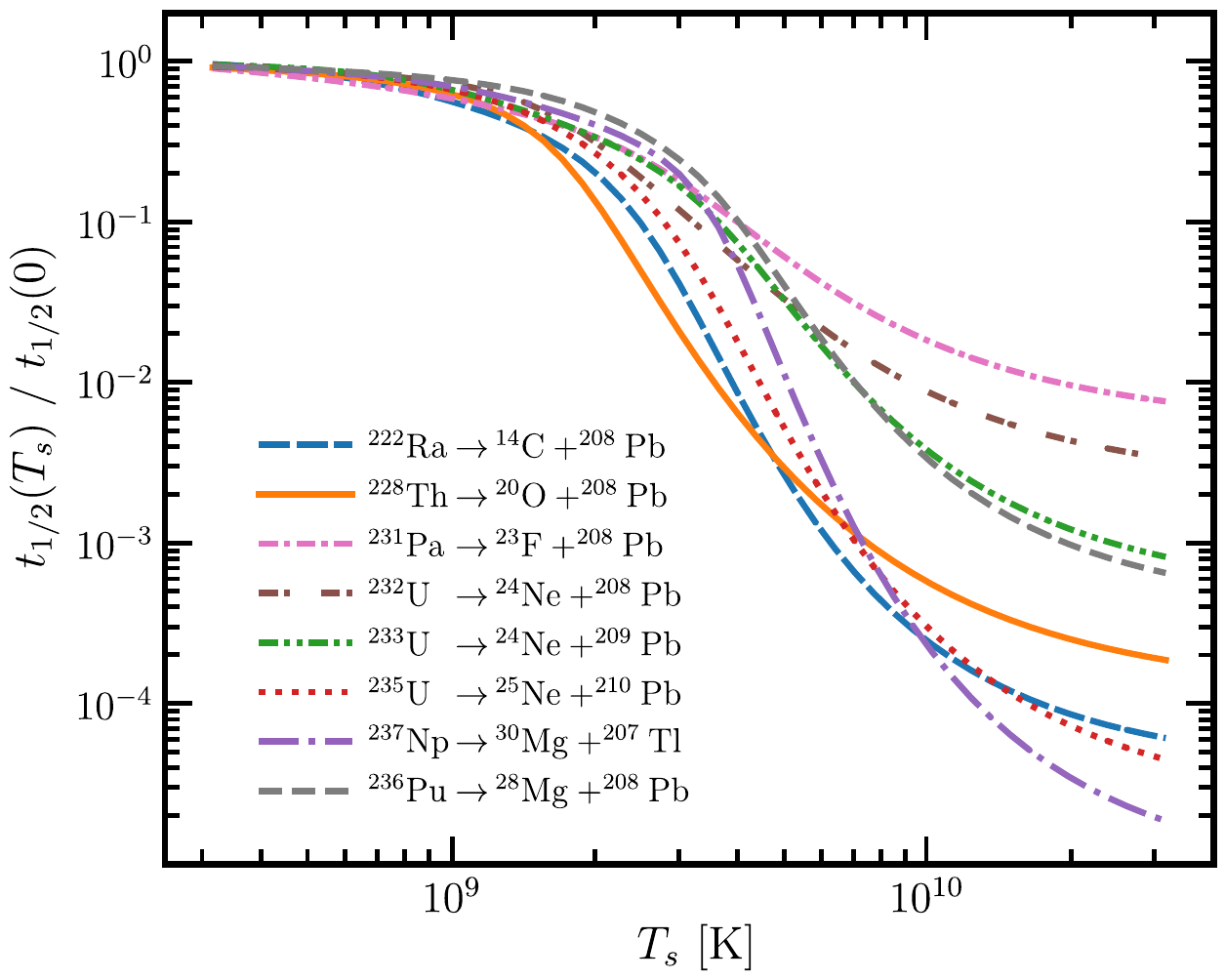}
    \caption{Ratio between the temperature-dependent cluster decay half-life $t_{1/2} (T_s)$ and that at zero temperature $t_{1/2} (0)$ for several representative temperatures scales.}
    \label{fig:halflife-temp}
\end{figure}

We examine the total enhancement of a given spontaneous emission of a light nucleus from excited states of a heavy nucleus (and resulting in an excited daughter nucleus) under conditions of finite temperature. For a given temperature of the environment $T_s$, the total cluster decay half-life is given by \cite{Ward-Fowler1980}
\begin{equation}\label{eq:T-half-life}
    t_{1/2} (T_s) = \left[\,\frac{1}{\mathcal{G}}\,\sum_{ij}\, \frac{g_{p_i}
    \exp\left(-E^*_{p_i}/k_BT_s\right)}{t_{1/2}\left(E_{p_i}^*,E_{d_j}^*\right)} \right]^{-1}\,,
\end{equation}
where $\mathcal{G}=\sum_{i} g_{p_i} \exp \left(-E^*_{p_i} / k_B T_s\right)$ is the standard partition function, and $g_{p_i}=(2J_{p_i}+1)$ is the statistical weight with $J_{p_i}$ being the spin of the parent state $p_i$. $k_B$ denotes Boltzmann's constant and $T_s$ is the surrounding temperature. $t_{1/2}\left(E_{p_i}^*,E_{d_j}^*\right)$ is calculated using (\ref{eq:half-life_temp_factorized}).

Figure (\ref{fig:halflife-temp}) displays the ratio between the temperature-dependent cluster decay half-life and that at zero temperature for several representative temperature scales occurring in different sites of r-process nucleosynthesis \cite{Cowan2021}. It is observed that each curve is flattened at both limits of temperature, namely $k_BT_s<<E_{p_{i}}^*$ and $k_BT_s>>E_{p_i}^*$. In the first case, when $k_BT_s<<E_{p_i}^*$, the exponential term acquires a very small value for all excited states. As a result, the major contribution is from the ground state decay for which $E_{p_0}^*=0$ resulting in the exponential term to be unity and hence removing the dependence on the temperature $T_s$. For the case when, $k_BT_s>>E_{p_i}^*$, the exponential term is close to unity for any excited state and once again the effects of the surrounding temperature are not significant. There is however a window between about 1 GK and 10 GK, in which the half-lives fall sharply with increasing temperature. This is to say that the decay rates are enhanced due to thermal excitations by few orders of magnitude in this temperature range. The latter can be a reason for the inclusion of the cluster decay channel in r-process nucleosynthesis calculations which do not consider this decay channel to be important, relying on the experimental information obtained from experiments performed in terrestrial laboratories.    

To summarize the above discussions, we can say that the above results show that the combined effect of considering both excited states in the initial and final states of decay and the temperature of the environment, is an enhancement in the probability of cluster decay. 

\section{Summary and Outlook}
Universal decay laws for the emission of light clusters are usually formulated by fitting a semi-empirical formula to the experimental half-lives. Here we extend such ideas to present an excitation energy dependent universal decay law (UDL) for the emission of light clusters from heavy nuclei. The excitation energy dependent half-lives evaluated within a standardly used theoretical model which reproduces the half-lives measured in the lab (i.e. from ground state to ground state) well, is used in order to fit the excitation energy-dependent parameters of the UDL. The excitation energy-dependence in the theoretical model enters through both the energy of the emitted cluster as an effective $Q$ value and the nuclear densities which are folded onto the elementary nucleon-nucleon interaction, thus giving us an excitation energy dependent double folding model. 

Given the findings of the present work, it seems important to consider the temperature and excitation energy dependent enhancement of decay rates in hot astrophysical environments. It would clearly be a formidable task to calculate the half-lives for hundreds of nuclei using a theoretical model and include it in the numerical codes for nucleosynthesis. A universal decay law such as the one presented in this work is therefore a first step in this direction and a combined excitation energy dependent UDL for different decay modes such as $\alpha$ decay, light cluster decay and fission could eventually be formulated for an appropriate modification of the nucleosynthesis calculations.   

\ack
D.F.R-G. thanks the Faculty of Science, Universidad de Los Andes, Colombia, for financial support through Grant No. INV-2021-126-2314. O. L. C. acknowledges support from NSERC and the Canadian Bureau for International Education.

\section*{References}
\bibliographystyle{unsrt}
\bibliography{bibliography}

\end{document}